%% file: main.tex
\documentclass[10pt,conference]{IEEEtran} 
\IEEEoverridecommandlockouts

\usepackage{xspace}
\usepackage{tcolorbox}
\usepackage{enumitem}

\usepackage{algorithm}
\usepackage{algpseudocode}
\algnewcommand\algorithmicforeach{\textbf{for each}}
\algdef{S}[FOR]{ForEach}[1]{\algorithmicforeach\ #1\ \algorithmicdo}
\algrenewcommand\alglinenumber[1]{\footnotesize #1}
\algrenewcommand\algorithmicrequire{\textbf{Input:}}
\algrenewcommand\algorithmicensure{\textbf{Output:}}

\usepackage{subcaption}
\usepackage{graphicx}

\usepackage{multirow}
\usepackage{booktabs}
\usepackage{tabularx}
\newcolumntype{C}[1]{>{\centering\arraybackslash}p{#1}}

\usepackage{amsmath}
\usepackage{mathtools} 

\usepackage[nice]{nicefrac}
\usepackage{xfrac}

\usepackage{booktabs}

\usepackage{amsthm} 

\usepackage{microtype}
\usepackage{url}
\usepackage{color, colortbl}
\usepackage{makecell}

\theoremstyle{acmdefinition}
\newtheorem{exmp}{\underline{Example}}[section]

\newcommand{\tool}{\textsc{Jarvis}\xspace}

\newcommand{\todo}[1]{\textcolor{black}{#1}}

\newcommand{\todonew}[1]{\textcolor{black}{#1}}
\newcommand{\todel}[1]{\textcolor{blue}{}}
\newcommand{\delnew}[1]{\textcolor{purple}{}}
\newcommand{\addnew}[1]{\textcolor{black}{#1}}

\def\BibTeX{{\rm B\kern-.05em{\sc i\kern-.025em b}\kern-.08em
    T\kern-.1667em\lower.7ex\hbox{E}\kern-.125emX}}

\begin{document}


\title{\tool: \addnew{Scalable and Precise Application-Centered Call Graph Construction for Python}}

\makeatletter
\newcommand{\linebreakand}{%
  \end{@IEEEauthorhalign}
  \hfill\mbox{}\par
  \mbox{}\hfill\begin{@IEEEauthorhalign}
}
\makeatother

\author{
  \IEEEauthorblockN{Kaifeng Huang}
  \IEEEauthorblockA{
    \textit{Tongji University}\\
    Shanghai, China
  }
  \and
  \IEEEauthorblockN{Yixuan Yan}
  \IEEEauthorblockA{
  \textit{Fudan University}\\
  Shanghai, China}
  \and
  \IEEEauthorblockN{Bihuan Chen}
  \IEEEauthorblockA{
    \textit{Fudan University}\\
    Shanghai, China}
    \and
  \IEEEauthorblockN{Yi Li}
  \IEEEauthorblockA{
    \textit{Fudan University}\\
    Shanghai, China}
  \and
  \IEEEauthorblockN{Zixin Tao}
  \IEEEauthorblockA{
    \textit{Fudan University}\\
  Shanghai, China}
  \and
  \IEEEauthorblockN{Xin Peng}
  \IEEEauthorblockA{
    \textit{Fudan University}\\
    Shanghai, China}
}

\maketitle

\begin{abstract}
\input{src/ch00-abstract}
\end{abstract}







\input{src/ch01-introduction}

\input{src/ch02-motivation}

\input{src/ch03-methodology-1}

\input{src/ch03-methodology-2}

\input{src/ch04-evaluation-1}
\input{src/ch04-evaluation-2}
\input{src/ch04-evaluation-3}

\input{src/ch05-related-work}

\input{src/ch07-conclusion}



\bibliographystyle{IEEEtran}
\bibliography{src/reference}


\end{document}

%% file: src/ch00-abstract.tex
Call graph construction is the foundation of inter-procedural~static~analysis. \textsc{PyCG} is the state-of-the-art approach for constructing call~graphs for Python programs. Unfortunately, \textsc{PyCG} does not scale to large~programs when adapted to whole-program analysis where \addnew{application and }dependent libraries are both analyzed. Moreover, \textsc{PyCG} is flow-insensitive \todo{and does not fully support Python's features}, hindering its accuracy.


To overcome these drawbacks, we propose a scalable \addnew{and precise} approach for constructing application-centered call graphs for Python programs,~and implement~it as a prototype tool \tool. \tool maintains a type graph (i.e., type relations~of program identifiers)~for each function in a program to allow \addnew{type inference}\delnew{reuse and improve scalability}. Taking one function as an input, \tool generates~the call graph on-the-fly, where flow-sensitive intra-procedural analysis and inter-procedural analysis are conducted in turn \addnew{and strong updates are conducted}. Our evaluation on a micro-benchmark of \todo{135} small Python programs~and~a~macro-benchmark of \todo{6} real-world Python applications has demonstrated that \tool can significantly improve \textsc{PyCG} by at least \todo{67\%} faster~in time, \todo{84\%} higher in precision, and at least \todonew{20\%} higher in recall.

%% file: src/ch01-introduction.tex

\section{Introduction}\label{sec:intro}

Python has become one of the most popular programming languages in recent years~\cite{python}. The prevalent adoption of Python in a variety of application domains calls for a great need for static analysis to ensure software quality. Call graph Construction (CGC) is the foundation of inter-procedural static analysis. It embraces a wide scope of static analysis tasks, e.g., security analysis~\cite{barros2015static, chess2004static, livshits2005finding, guarnieri2009gatekeeper}, dependency management~\cite{hejderup2018software, wang2022Insight, huang2022empirical} and software debloating~\cite{quach2018debloating, Bruce2020}.

\todonew{\textbf{Problem.}} It is challenging to generate a precise and sound call graph for \todonew{interpreted languages~\cite{jensen2009type, nagy2021unambiguity, nielsen2021modular}}. \todel{Python has dynamic language features as any interpreted language does, which makes the analysis more complicated compared with compiled languages~\cite{jensen2009type, nagy2021unambiguity, nielsen2021modular}.} For example, dynamically typed variables demand the analysis to undertake~a~precise inter-procedural points-to analysis to resolve the type of variables. 
\todonew{In response to the increasing demand for Python, several CGC approaches have been recently proposed, i.e., \textsc{Pyan}~\cite{pyan}, \textsc{Depends}~\cite{depends} and \textsc{PyCG}~\cite{salis2021pycg}.}
\todonew{Among these CGC tools for Python}\todel{Specifically}, \textsc{PyCG} is the state-of-the-art, which achieves the best \todonew{accuracy}\todel{precision, recall}, time and memory performance~\cite{salis2021pycg}.~It~conducts a flow-insensitive inter-procedural points-to analysis using a \todonew{worklist}\todel{fixed-point} algorithm~\cite{rayside2002generic, lhotak2003scaling}. This algorithm takes unfixed iterations~to~update an \textit{assignment graph} (i.e., points-to relations~between program identifiers) until it is unchanged.~After~the~assignment graph~for~the whole program is constructed, the call graph is generated.

\addnew{\textbf{Limitations.} Whole-program CGC is important~to support security analysis, dependency management and~software debloating.
However, \textsc{PyCG} encounters a scalability challenge, particularly for large applications which include hundreds of dependent libraries. 
This issue is reflected in our experiment, where \textsc{PyCG} encounters out-of-memory and recursion error in two projects and exceeds the time limit in the rest four projects, as illustrated in Table \ref{table:intro}.
Moreover, \textsc{PyCG} conducts a flow-insensitive analysis which ignores control flows and over-approximately computes points-to relations. This over-approximation introduces false positives. In addition, \textsc{PyCG} does not fully support Python's language features (e.g., context managers and built-in types), which introduces false negatives. Such false positives and false negatives\todel{ introduced by \textsc{PyCG}} would amplify the inaccuracy in longer call chains.}


\begin{table}[!t]
  \centering
  \footnotesize
  \caption{Scalability Results of \textsc{PyCG} on Our Benchmark (Time Limit is 24 Hours, Memory Limit is 8GB)}\label{table:intro}
  \vspace{-5pt}
  \begin{tabular}{m{0.5cm}m{0.3cm}m{0.45cm}m{0.5cm}m{2.5cm}m{2.0cm}}   
  \noalign{\hrule height 1pt}
  Project & LOC (A.)  & LOC (W.) &  Lib. & Time (hours) & Memory (GB) \\
  \noalign{\hrule height 1pt}
  P1 & 5.0k & 120.5k  & 201 & \cellcolor{lightgray}  Exceeding time limit &2.3G \\
  P2 & 4.8k & 108.4k   & 191   &  \cellcolor{lightgray} Exceeding time limit &5.7G \\
  P3 & 0.5k & 515.3k   & 237   &   4h+& \cellcolor{lightgray} Recursion error \\
  P4 & 2.9k & 109.3k   & 190   &  \cellcolor{lightgray} Exceeding time limit &5.1G  \\
  P5 & 1.1k & 200.4k   & 192   & 0.5h+& \cellcolor{lightgray} Out of memory\\
  P6 & 2.5k & 141.4k   & 202  & \cellcolor{lightgray} Exceeding time limit &4.6G \\
  \noalign{\hrule height 1pt}
  \end{tabular}
\end{table}

\begin{figure*}[!t]
  \centering
  \includegraphics[scale=0.24]{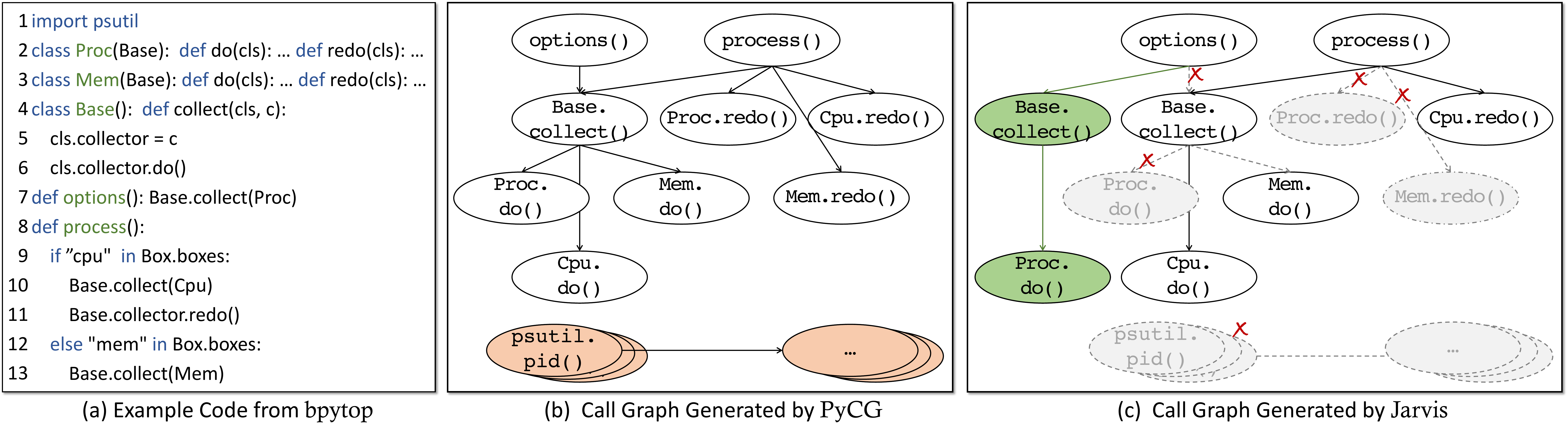}
  \vspace{-13pt}
  \caption{Motivating Example Code and Call Graphs generated by\textsc{PyCG} and \tool}
  \label{fig:example_code} 
\end{figure*}

\addnew{\textbf{Approach.} We propose a scalable and precise call graph~construction approach for Python, and implement it as a prototype~tool \tool. \tool has three characteristics~which are different from \textsc{PyCG} and existing works. First, \textit{\tool is \textbf{application-centered};} i.e., \tool only constructs call relations and reachable callees from the application functions to the invocated library functions which we refer to as application-centered. In some cases such as tracing vulnerable library functions from the application or debloating library functions by application's usage, application-centered CGC is more efficient than whole-program CGC. As can be seen from Table \ref{table:intro}, the projects depend on more than 190 libraries, and the scale of the whole program (W.) is significantly larger than the application (A.) regarding the LOC. However, existing whole-program CGC methods could not be effortlessly migrated to application-centered CGC. Using application-centered CGC, \tool is scalable to applications with numerous depending libraries. }

\todonew{Second, \textit{\tool conducts flow-sensitive type inference.} \tool maintains a local type graph for each function in a program. Such a design allows \tool to \addnew{create~a copy of type graph when control flow diverges~in~a~function and merging type graphs when control flow converges, and reuse the type graph~of~a~function without reevaluating the function~at~each call~site.} This allows \tool to be flow-sensitive and \addnew{able to conduct strong updates} if there are no loops, which would result in precise type inference.~Third, \textit{\todo{\tool supports Python's language features more comprehensively}}, which improves its recall. Based on these three characteristics, by constructing the call graph on-the-fly, \tool is scalable to large programs.}


\textbf{Evaluation.} We evaluate the efficiency and effectiveness of \tool~on~a~micro-benchmark of \todo{135} small Python programs \todel{(covering a wide range of language features) }and a macro-benchmark of \todo{6} real-world Python applications. For efficiency, \addnew{\tool takes an average of 8.16 seconds and 227 MB in application-centered CGC.} \tool improves \textsc{PyCG} by at least \todo{67\%} faster in time in the scenario of whole-program CGC.~For effectiveness, \tool improves \textsc{PyCG} by \todo{84\%} in precision~and~at least \todonew{20\%} in recall in application-centered CGC. \todo{We also demonstrate the application of \tool in dependency management.}
 
In summary, this work makes the following contributions.
\begin{itemize}[leftmargin=*]
\item We proposed \tool to scalably construct application-centered call graphs for Python~programs via type inference.
\item We conducted experiments on two benchmarks to demonstrate~the improved efficiency and effectiveness over the state-of-the-art.
\item \todonew{We conducted a case study to demonstrate the application of \tool in dependency management.}
\end{itemize}

%% file: src/ch02-motivation.tex

\section{Motivating Example}\label{sec:motivation}

We discuss the drawbacks of \textsc{PyCG} and explain the insights of \tool using a motivating example. 

\addnew{Fig. \ref{fig:example_code}(a) presents a snippet from \texttt{bpytop.py} in the \texttt{bpytop} project~\cite{bpytop}. It is a resource monitor that shows usage and stats for cpu, memory, processes, etc. We have made simplifications to the snippet for clearer understanding. The snippet imports \texttt{psutil} which is a dependent library. \texttt{Proc}, \texttt{Net} and \texttt{Mem} are the subclasses of \texttt{Base} where they all implement the \texttt{do()} and \texttt{redo()} methods. The function \texttt{process()} and \texttt{options()} are two entrances which would invoke the \texttt{do()} and \texttt{redo()} methods from \texttt{Proc},~\texttt{Net}~or~\texttt{Mem}.}

\addnew{We choose \textsc{PyCG}, which is the state-of-the-art CGC method for Python. \textsc{PyCG} adopts the worklist algorithm~\cite{rayside2002generic, lhotak2003scaling} where iterations are unfixed until the global assignment graph remains unchanged. Fig. \ref{fig:example_code}(b) presents the call graph generated by \textsc{PyCG}. It suffers~from~two~issues.}

\begin{itemize}[leftmargin=*]
    \item \addnew{\textbf{Scalability issue.} \textsc{PyCG} treats the entire program equally without considering its hierarchy within the application and external dependencies. However, these dependencies are common due to open-source software supply chains. In this context, users of the call graph are interested in understanding how the application interacts with its dependencies, rather than focusing on outliers. The function \texttt{psutil.pid()} and its callees, filled in red in Fig. \ref{fig:example_code}(b), denote that those outliers are not connected with the functions from the application. The computation for the outliers may be unnecessary because they are not reachable from the entry file \texttt{bpytop.py}. Yet it can be computationally expensive, especially in the case of library bloating \cite{porter2020blankit, soto2021comprehensive, xin2022studying}.}
    \item \addnew{\textbf{Preciseness issue.} \textsc{PyCG}~generates~call~relations~connecting \texttt{options()}~to~\texttt{Base.collect()},~and \texttt{process()}~to~\texttt{Base.collect()}.~The~invocation~of \texttt{Base.collect()} accepts \texttt{Proc} (Line 7), \texttt{CPU} (Line 10) and \texttt{Mem} (Line 13) as parameters. Due to its flow-insensitive nature, \textsc{PyCG} would create points-to relations from \texttt{Base.collector} to \texttt{Proc}, \texttt{CPU} and \texttt{Mem} overapproximately. Consequently, when evaluating \texttt{Base.collector.redo()} (Line 11) in the next iterations, the points-to relations would create spurious call relations from \texttt{process()} to \texttt{Proc.redo()},~\texttt{Mem.redo()}, except for the only correct call relation from \texttt{process()} to \texttt{Cpu.redo()}. Additionally, \textsc{PyCG} reports two false call relations from \texttt{Base.collect()} to \texttt{Mem.do()} and \texttt{Cpu.do()} when calling \texttt{options()}, which is due to that \textsc{PyCG} does not distinguish different calling contexts in whole-program analysis. Similarly, \textsc{PyCG} reports one false call relation from \texttt{process()} to \texttt{Proc.do()} as well.}
    
\end{itemize}

\addnew{We demonstrate how \tool can address these challenges through application-centered CGC. 
The key challenge in generating call graphs is finding a balance between precision and scalability.}
\addnew{Fig. \ref{fig:example_code}(c) shows the call graph generated by \tool. We use \texttt{options()} and \texttt{process()} as queries to generate application-centered call graphs. The dotted lines and grey nodes denote that those present in the original call graph are absent in ours. }\addnew{Overall, \tool differs from prior work in three aspects.}

\begin{itemize}[leftmargin=*]
    \item \addnew{\textbf{Application-centered call graph construction.} \tool takes the entrance function as an input, and generates the call graph on-the-fly, which can avoid over-approximation to some extent and be adapted to application scenarios where call graphs need to be constructed promptly. In Fig. \ref{fig:example_code}(c), the red ovals turned light grey. Those outliers are not reachable from the entrance function, thereby would not be evaluated. Considering the scale of bloating libraries, it can significantly reduce the computational cost. Besides, in Fig. \ref{fig:example_code}(c), we turned \texttt{Proc.do()} to light grey, and created the calling chain from \texttt{options()} to \texttt{Base.collect()}, \texttt{Proc.do()} denoted in green. By adopting application-centered CGC, we isolate the points-to relations of each entrance function, thereby spurious call relations are avoided.}
    \item \addnew{\textbf{Flow sensitive type inference.} Existing CGC techniques are not flow-sensitive, and are not capable of handling strong updates. In Fig. \ref{fig:example_code}(c), we turned \texttt{Proc.redo()} and \texttt{Mem.redo()} to light grey. The two spurious call relations in the original call graph are caused by flow-insensitiveness when evaluating \texttt{Base.collector.redo()} (Line 11). \tool creates and maintains function type graphs to enable flow-sensitive analysis. It can precisely track the data flow and control flow so that strong updates of types can be performed, which can avoid false positives and improve the preciseness~of~call~graphs.}
    \item \addnew{\textbf{New support for Python language features.} Existing CGC techniques do not support Python's language features comprehensively. We provide new support on Pythons' language features such as Python context managers, and provide enhancements on existing features.}
\end{itemize}

%% file: src/ch03-methodology-1.tex

\section{Approach}\label{sec:approach}

We first explain  give notions and domains of our analysis. Then, we describe the overview of \tool. Finally, we elaborate \tool in detail.



\subsection{Notions and Domains}\label{sec:approach:notion}

Our analysis works on the AST representation of Python programs using the built-in \textit{ast} library in Python. 
Fig.~\ref{fig:notion} shows the basic notions used in our analysis. Specifically, each identifier type $t$ is defined as a tuple $\langle \tau, \phi, n \rangle$, where $\tau$ denotes the identifier element, $\phi$ denotes the identifier namespace, and $n$ denotes the identifier name. $\tau$ can~be one of the five types, i.e., module in the application (\textbf{mod}), module in~external dependent libraries (\textbf{ext\_mod}), class (\textbf{cls}), function~(\textbf{func}), and variable (\textbf{var}). Each expression $e$ can be of various types, and a full list of expressions can be found at the official Python website~\cite{pythonast}. Then, we introduce the six domains that are maintained in our analysis.

\begin{figure}[!htb]
    \begin{align*}
        d \in Type ::&= \langle \tau, \phi, n \rangle \\
        \tau \in IdentifierElemt ::&= \textbf{mod} \mid \textbf{ext\_mod} \mid \textbf{cls} \mid \textbf{func} \mid \textbf{var} \\
        n  \in IdentifierName::&= the~set~of~program~identifier \\& names \\
        \phi  \in Namespace ::&= (n, \tau)^* \\
        e \in Expr ::&= the~set~of~expressions
    \end{align*}
    \vspace{-20pt}
    \caption{Notions of Our Analysis}
    \label{fig:notion} 
\end{figure}

\textit{\textbf{Function Type Graph}} ($\mathcal{FTG}$). \textsc{PyCG} maintains the assignment graph (AG) globally, which consists of assignment~relations between program identifiers program-wide.
To avoid unnecessary analysis and enable flow-sensitiveness in application-centered CGC, we maintain function type graphs (FTG).
\addnew{A function type graph maintains type relations (i.e., classes, methods, functions) of a function, which is different from the program-level assignment graph in \textsc{PyCG}.~It~is~designed at the function level to allow reuse and improve scalability.} Formally, a function assignment graph is denoted as a 3-tuple $\langle Type, \todo{Expr}, R \rangle$, where $Type$ denotes identifier types in a function, $Expr$ denotes expressions in a function, and~$R$ denotes type relations between identifiers. Each type relation $rls \in R$ is denoted as a 3-tuple $\langle t_1, t_2, e\rangle$ (or $t_1 \xrightarrow{e} t_2$),~where~$t_1$, $t_2$ $\in$ $Types$, and $e \in Expr$ denotes the evaluated expression that~results in~$rls$. Here $e$ facilitates flow-sensitive analysis \todo{(see Sec.~\ref{sec:approach:intra:computeFTGOutput}).}


Hereafter, we use $\mathcal{FTG}_{in}$ to denote the initial FTG before~the~evaluation of the first expression in a function, which contains type relations about parameter variables that are passed from its caller. We use $\mathcal{FTG}_{e}$ to denote the intermediate FTG after the expression $e$ is evaluated. We use $\mathcal{FTG}_{R}$ to denote the final FTG~after~all~expressions are evaluated. We use $\mathcal{FTG}_{out}$ to denote the output~FTG which contains the final type relations about parameter variables that will be passed back to its caller.


\textit{\textbf{Control Flow Graph}} ($\mathcal{CFG}$). The control flow graph of a function maintains the control dependencies between expressions. It is denoted as a 4-tuple $\langle \todo{Expr}, \todo{Ctrl}, \todo{e_{en}}, E_{r}\rangle$, where $Expr$ denotes~expressions in a function, $Ctrl$ denotes control flows in a function, 
$e_{en}$ denotes the entry expression, and $E_{r}$ denotes the return expressions (either explicit returns or implicit returns). Each control~flow $ctrl \in$ $Ctrl$ is denoted as a 2-tuple $\langle e_1, e_2\rangle$, representing the control flow from expression $e_1$ to expression $e_2$. Notice that we add a virtual dummy expression $e_{dum}$ that all return expressions flow to.



\textit{\textbf{Function Summary}} ($\mathcal{F}$). The function summary contains a set of functions that are visited in our analysis. Each function $f \in \mathcal{F}$ is denoted as a 2-tuple $\langle t, P \rangle$, where $t \in Type$ denotes the type of $f$, and $P$ denotes the set of parameter names of $f$. 

\textit{\textbf{Class Summary}} ($\mathcal{C}$). The class summary is denoted as a 2-tuple $\langle \todo{Hier}, \todo{Incl}\rangle$, where $Hier$ denotes class hierarchy (i.e., inheritance relations between classes), and $Incl$ denotes the inclusion relations from class to its included function types. Each entry in $Hier$ is denoted as a 2-tuple $\langle t_{clsb},$ $t_{clss} \rangle$, where $t_{clsb}$ and $t_{clss} \in$ $Type$ denote the base calss and sub~class, respectively. Each entry in $Incl$ is denoted~as a 2-tuple $\langle t_c, t_f \rangle$, where $t_c \in Type$ denotes a class, and $t_f \in Type$ denotes a function type in the class. 

\textit{\textbf{Import Summary}} ($\mathcal{I}$). The import summary contains type relations from the importing type to the imported type. Each entry~in~$\mathcal{I}$~is~denoted as a 3-tuple $\langle \todo{t_s}, t_t, e\rangle$, where $t_s \in Type$ denotes the importing type, $t_t \in Type$ denotes the imported type, and $e$ denotes the import expression. 
  
\textit{\textbf{Call Graph}} ($\mathcal{CG}$). The call graph contains call relations~in~a~program. It is denoted as a 2-tuple $\langle V, E\rangle$, where each entry in $E$~is~denoted as a 2-tuple $\langle f_{er}, f_{ee} \rangle$, representing a call relation from the caller function $f_{er} \in V \subseteq Type$ to the callee function $f_{ee} \in V$.

\subsection{Overview of \tool}\label{sec:approach:overview}

\begin{figure}[!t]
    \centering
    \includegraphics[scale=0.334]{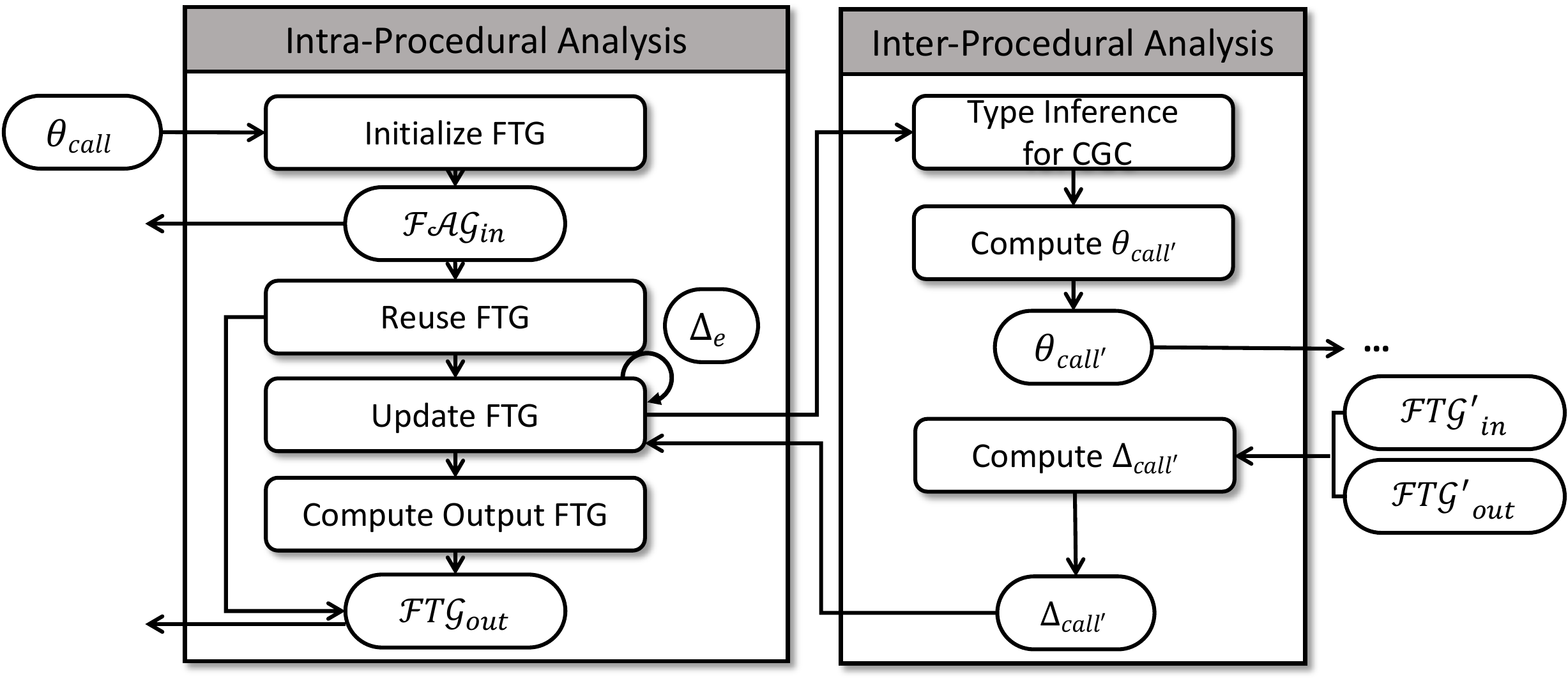}
    \vspace{-5pt}
    \caption{Approach Overview of \tool}
    \label{fig:tool_overview} 
\end{figure}

The overview of \tool is shown in Fig. \ref{fig:tool_overview}. Given an entry~function, it first runs intra-procedural analysis on this function, and then runs inter-procedural analysis and intra-procedural analysis in turn.

\textbf{Intra-Procedural Analysis.} When running intra-procedural analysis on a function $f$, \tool takes as input $\theta_{call}$, which denotes the type relations about argument variables passed from the call expression $call$ that invokes $f$. For an entry function,~its~$\theta_{call}$~is empty by default. \tool takes four steps to complete intra-procedural analysis. First, \tool initializes $\mathcal{FTG}_{in}$ before evaluating the function's expressions. Then, \tool reuses $\mathcal{FTG}_{out}$ if the function~has been visited before and the type relations about argument variables are the same as before. Otherwise, \tool updates the FTG.~It applies transfer functions to evaluate each expression $e$ in the function, and adds the evaluated result in $\Delta_e$ to obtain $\mathcal{FTG}_{e}$. Specifically, if $e$ is a call expression $call^\prime$, \tool runs an inter-procedural analysis to obtain its evaluated result $\Delta_{call^\prime}$. Finally, after all expressions are evaluated, \tool computes $\mathcal{FTG}_{out}$, and passes~$\mathcal{FTG}_{in}$ and $\mathcal{FTG}_{out}$ back to the inter-procedural~analysis~on~$call$.

\textbf{Inter-Procedural Analysis.} \tool starts inter-procedural analysis when a call expression $call^\prime$ in $f$ is resolved into a callee function $f^\prime$. Specifically, \tool first 
 creates a call relation $\langle f, f^\prime \rangle$ and updates it into the call graph. Then,~\tool computes type relations about argument variables in $call^\prime$ (i.e., $\theta_{call^\prime}$). Next, \tool runs intra-procedural analysis on $f^\prime$ using $\theta^{\prime}_{call}$, and passes $\mathcal{FTG}^{\prime}_{in}$ and $\mathcal{FTG}^{\prime}_{out}$ back. Finally, \tool~computes the changed type relations about argument variables~(i.e., $\Delta_{call^\prime}$) from $\mathcal{FTG}^{\prime}_{in}$ to $\mathcal{FTG}^{\prime}_{out}$, and passes $\Delta_{call^\prime}$ back to the intra-procedural analysis on $f$ in order to reflect the evaluated results from $call^\prime$.
 

After our intra-procedural analysis and inter-procedural analysis finish, we can directly obtain the call graph because it is constructed on the fly in our inter-procedural analysis.

%% file: src/ch03-methodology-2.tex

\begin{algorithm}[!t]
    \footnotesize
    \caption{Intra-Procedural Analysis}\label{alg1}
    \begin{algorithmic}[1]
        \Require $f$, {$\theta_{call}$} 
        \Ensure $\mathcal{FTG}^f_{in}$, {$\mathcal{FTG}^f_{out}$}
    \Function{intra\_analysis}{}
    \State $g_{in}$ = computeParamType($f.P$, $\theta_{call}$)
    \If{$not$ isVisited($m_f$)}
        \State $\mathcal{F} , \mathcal{C}, \mathcal{I}$ $\leftarrow$ previsitModule($m_f$)
    \EndIf
    \State $g_{in} \leftarrow \mathcal{I}_{m_f}$

    \If{isVisited($f$) and isEqual($g_{in}$, $\mathcal{FTG}^f_{in}$)} 
        \State \Return $\mathcal{FTG}^f_{in}$, $\mathcal{FTG}^f_{out}$
    \EndIf

    \State $\mathcal{FTG}^f_{in}$ = $g_{in}$
    \ForEach {$e \in$ preOrder(AST of $f$)}
        \ForEach {$e.p \in$ parents of $e$ and outDegree($e.p$) $> 1$}
             \State $\mathcal{FTG}^f_{e.p}$  = copy($\mathcal{FTG}^f_{e.p}$)
        \EndFor
        \If{$e$ has one parent}
            \State $\Delta_e$ = applyTransferRule($f$, $e$, $\mathcal{FTG}^f_{e.p}$)
            \State $\mathcal{FTG}^f_{e}$ = updateFTG($\mathcal{FTG}^f_{e.p}$, $\Delta_e$, $e$)
        \EndIf
        \If{$e$ has more than one parent}
            \State $g_{merge} = \bigcup_{e.p \in parents~of~e}  \mathcal{FTG}^f_{e.p}$
            \State $\Delta_{e}$ = applyTransferRule$(f, e, g_{merge})$
            \State $\mathcal{FTG}^f_{e}$ = updateFTG$(g_{merge}, \Delta_e, e)$
        \EndIf  
    \EndFor
    \State $ \mathcal{FTG}^f_{R}= \bigcup_{e \in \mathcal{CFG}_f.E_{r}} \mathcal{FTG}^f_{e} $
    \State setVisited($f$)
    \State $\mathcal{FTG}^f_{out} $ = computeOutputFTG$(\mathcal{FTG}^f_{in}, \mathcal{FTG}^f_{R})$
    \State \Return $\mathcal{FTG}^f_{in}$, $\mathcal{FTG}^f_{out}$
    \EndFunction
    \end{algorithmic}
\end{algorithm}

\subsection{Intra-Procedural Analysis}\label{sec:approach:intra}

The algorithm of our intra-procedural analysis is presented in Alg.~\ref{alg1}. It mainly consists of four steps, i.e., \textit{Initialize FTG}, \textit{Reuse FTG}, \textit{Update FTG}, and \textit{Compute Output FTG}.  

\subsubsection{\textbf{Initialize FTG}}\label{sec:approach:intra:init} 

Given the type relations $\theta_{call}$ about argument variables in the call expression $call$ that invokes function~$f$, this step initializes $\mathcal{FTG}^f_{in}$ (Line 2--6 in Alg.~\ref{alg1}), i.e., the initial FTG of $f$ before evaluating the first expression in $f$. First, it computes the type relations about parameter variables (i.e., $f.P$) according to $\theta_{call}$ based on the mapping between parameters and passed arguments, and puts the result to a temporary FTG $g_{in}$ (Line 2).~Then, if the module $m_f$ where $f$ locates (which can be derived from $f.d.\phi$) has not been visited,~it~updates $\mathcal{F}$, $\mathcal{C}$ and $\mathcal{I}$ by parsing the code~of~$m_f$ (Line 3--5). Finally, it adds the global type relations resulting from import expressions to $g_{in}$ (Line~6).~Now~$g_{in}$~is the~initial FTG of $f$, and will~be~assigned~to~$\mathcal{FTG}^f_{in}$~(Line 10).

\subsubsection{\textbf{Reuse FTG}}\label{sec:approach:intra:reuse} 

If the function has been visited before, the FTG of $f$ has been constructed, which provides the opportunity to reuse the FTG and improve scalability. To this end, this step first determines whether $f$ has been visited before (Line 7). If yes, it further determines whether $g_{in}$ equals~to~$\mathcal{FTG}_{in}^{f}$ that is constructed~in~the previous visit (Line 7). If yes (meaning that the FTG can be reused), it directly returns the previously constructed $\mathcal{FTG}_{in}^f$ and $\mathcal{FTG}_{out}^f$ (i.e., the final type relations about parameter variables) (Line 8). Otherwise, it goes to the next step to build the FTG.


\begin{figure}[!t]
    \raggedleft
    \footnotesize
        \resizebox{.98\linewidth}{!}{
          \begin{minipage}{\linewidth}
            \vspace*{-10pt}
    \begin{align*}
        &\text{Import:}~from~m~^\prime~import~x, import~m^\prime \\
        &\frac{\begin{matrix}
            t_1=new\_type(m,~x), t_2=new\_type(m^\prime,~x), t_3=t_m, t_4=new\_type(m^\prime)
            \end{matrix}
            }{ \Delta_{e} \leftarrow \langle t_1, t_2, e\rangle, \Delta_{e} \leftarrow \langle t_3, t_4, e\rangle}\\
        &\text{Assign:}~x=y \\ 
        &\frac{t_1=new\_type(x), t_2=new\_type(y)} { \Delta_{e} \leftarrow \langle t_1, t_2, e\rangle}  \\
        &\text{Store:}~x.field~=~y\\
        &\frac{t_i \in points(x), t_2 = new\_type(y)}{\Delta_{e} \leftarrow \langle t_i.field, t_2, e\rangle}\\
        &\text{Load:}~y~=~x.field\\
        &\frac{t_1 \in new\_type(y), t_j \in points(x)}{\Delta_{e} \leftarrow \langle t_1, t_j.field,~e \rangle}  \\
        &\text{New:}~y~=~x(...) \\
        &\frac{t_1=new\_type(y), t_2=new\_type(x)}{
            \begin{matrix}
            \Delta_{e}~\leftarrow~\textsc{inter\_analysis}(f,~e,~\mathcal{FTG}^f_{e.p}), \Delta_{e}~\leftarrow~\langle~t_1,~t_2,~e~\rangle\\
            \end{matrix}
            } \\
        &\text{Call:}~a=x.m(...) \\
        &\frac{
            \begin{matrix}
            t_1=new\_type(x), t_2=new\_type(t_1.m), t_3=new\_type(a)
            \end{matrix}
            }
            {\begin{matrix}
                \Delta_{call}~\leftarrow~\textsc{inter\_analysis}(f,~e,~\mathcal{FTG}^f_{e.p}), \Delta_{call}~\leftarrow~\langle~t_3,~t_2.\textit{ret},~e\rangle
            \end{matrix}
                }\\
        &\text{Return:}~def~m^\prime ...~return~x\\
        &\frac{t_1=new\_type(m^\prime), t_2=new\_type(x)}{\Delta_{e} \leftarrow \langle t_1.\textit{ret}, t_2, e \rangle}
    \end{align*}
\end{minipage}
}
 \vspace{-3pt}
    \caption{Transfer Rules of Our Analysis}
    \label{fig:rules} 
\end{figure}

\subsubsection{\textbf{Update FTG}}\label{sec:approach:intra:updFTG}

Given $\mathcal{FTG}_{in}^f$, this step iterates each expression by a preorder traversal on the AST of $f$ (Line 11--24 in Alg.~\ref{alg1}). The control flow graph~(i.e.,~$\mathcal{CFG}_f$) and the FTG are updated using each expression's evaluated result in each iteration. $\mathcal{CFG}_f$~is~used~to enable flow-sensitive analysis, but we omit its straightforward construction detail. In each iteration,  three steps depend on the positive of the evaluated expression $e$~in~$\mathcal{CFG}_f$.

First, for each of $e$'s parent expressions (denoted as $e.p$), if the out-degree of $e.p$ in $\mathcal{CFG}_f$ is larger than 1 (Line 12), it means~that the control~flow diverges with $e$ as the first expression on the diverged flow. To enable flow-sensitive analysis, it creates a copy of $\mathcal{FTG}^f_{e.p}$, i.e., the FTG after $e.p$ is evaluated, for further update (Line 13). 

Second, if $e$ has one parent in $\mathcal{CFG}_f$, the resulting type relations $\Delta_{e}$ from evaluating $e$ is strong-updated to $\mathcal{FTG}^f_{e.p}$ for producing $\mathcal{FTG}^f_{e}$ (Line 15--18). The evaluation applies a transfer function, which is comprised of a list of transfer rules with respect to different expressions. Part of the transfer rules are reported in Fig. \ref{fig:rules}, and a full list is available on our website. Each transfer rule generates new type relations $\Delta_{e}$. Notice that if $e$ is a call expression, it runs inter-procedural analysis to compute $\Delta_{e}$ (see. Sec. \ref{sec:approach:inter:delta}). After~$\Delta_{e}$~is computed, $\mathcal{FTG}^f_{e}$ is computed by adding $\Delta_{e}$ to $\mathcal{FTG}^f_{e.p}$. 

Third, if $e$ has more than one parent, for all of $e$'s parent~expressions, it merges their $\mathcal{FTG}^f_{e.p}$ into a new FTG $g_{merge}$ (Line~20),~representing that the control flow converges. Then, it evaluates $e$ and adds newly generated type relations $\Delta_{e}$ to $g_{merge}$ (Line 21--22).

After all iterations finish, for each of the return expressions,~i.e., $e \in \mathcal{CFG}_f.E_r$, there exists $\mathcal{FTG}^f_e$. Our analysis proceeds to merge them into $\mathcal{FTG}^f_{R}$, i.e., the final FTG after all expressions in $f$ have been evaluated (Line 25).

\subsubsection{\textbf{Compute Output FTG}}\label{sec:approach:intra:computeFTGOutput}

Given $\mathcal{FTG}^f_{in}$ and $\mathcal{FTG}^f_{R}$, this step computes $\mathcal{FTG}^f_{out}$ (Line 27). Different from $\mathcal{FTG}^f_{R}$ which records possible type relations across all control flows, $\mathcal{FTG}^f_{out}$ is a subset of $\mathcal{FTG}^f_{R}$ where type relations about temporary variables are discarded and the rest of type relations are still in effect after the call to $f$ is returned. In particular, for each type relation $t_1 \xrightarrow{e} t_2$ $\in$ $\mathcal{FTG}^f_{in}$, it first obtains $t_1$ from $\mathcal{FTG}^f_{in}$, and then computes types that $t_1$ finally is in $\mathcal{FTG}^f_{R}$.~Notice that it also computes types that the fields of $t_1$ finally~is. Concerning that $t_1$ may refer to multiple types, it compares the order of evaluated expression for each type relation using $\mathcal{CFG}_f$ in order to select the latest result as $\mathcal{FTG}^f_{out}$.

\begin{exmp}
\addnew{Using $\mathcal{CFG}_f$, we can learn that \texttt{Base.collector} is only the type of \texttt{Cpu} after the expression in Fig. \ref{fig:example_code}(a) at Line 10 is evaluated. $\mathcal{FTG}^{process()}_{R}$ contains two type relations from \texttt{process()} after \texttt{process()} in Fig. \ref{fig:example_code}(a) is evaluated, i.e., \texttt{Base.collector} $\xrightarrow{bpytop:10}$ \texttt{Cpu}, and \texttt{Base.collector} $\xrightarrow{bpytop:13}$\texttt{Mem}.}
\end{exmp}

\subsection{Inter-Procedural Analysis}\label{sec:approach:inter}

The algorithm of our inter-procedural analysis is shown in Alg.~\ref{alg2}. It has three steps, i.e., \textit{Type Inference for CGC}, \textit{Compute $\theta_{call^{\prime}}$}, and \textit{Compute}~$\Delta_{call^{\prime}}$.

\begin{algorithm}[!t]
\footnotesize
    \caption{Inter-Procedural Analysis}\label{alg2}
    \begin{algorithmic}[1]
        \Require ${f}$, ${call^\prime}$, $\mathcal{FTG}^f_{call^{\prime}.p}$
        \Ensure \textit{$\Delta_{call^{\prime}}$}
    \Function{inter\_analysis}{}
        \State $f^{\prime}$ = updateCG$(call^\prime, \mathcal{FTG}^f_{call^{\prime}.p}, \mathcal{I}, \mathcal{F}, \mathcal{C})$
        \State $\theta_{call^{\prime}}$ = compute$\theta_{call^{\prime}}$($\mathcal{FTG}^f_{call^{\prime}.p}$, $call^{\prime}$)
        \State $\mathcal{FTG}^{f^\prime}_{in}, \mathcal{FTG}^{f^\prime}_{out}$ = \textsc{intra\_analysis}$(f^{\prime}, \theta_{call^{\prime}})$
        \State $\Delta_{call^{\prime}} $= compute$\Delta_{call^{\prime}}(\mathcal{FTG}^{f^{\prime}}_{in}, \mathcal{FTG}^{f^{\prime}}_{out})$
        \State \Return $\Delta_{call^{\prime}} $
    \EndFunction
    \end{algorithmic}
\end{algorithm}

\subsubsection{\textbf{\addnew{Type Inference for CGC}}}\label{sec:approach:inter:gencg} 

\tool generates call relations during our inter-procedural analysis (Line 2 in Alg.~\ref{alg2}).  
Given inputs as the call~expression $call^\prime$ at the caller function $f$ as well as the FTG after evaluating $call^\prime$'s parent expression $\mathcal{FTG}^f_{call^{\prime}.p}$, this step resolves the callee function $f^\prime$ \addnew{using type inference from $\mathcal{FTG}^f_{call^{\prime}.p}$}, and adds the call relation $\langle f, f^\prime \rangle$~to~$\mathcal{CG}$. 

Specifically, $call^\prime$ can be in the form of \texttt{a.b(...)} or \texttt{b(...)}. For the form of \texttt{a.b(...)}, it first searches $\mathcal{FTG}^f_{call^{\prime}.p}$ for the~class~types that are pointed to by the invoking variable (e.g., \texttt{a}) of $call^\prime$. Then, for each searched class type $t_{cls}$, it checks whether~there exists a function type $t_{b}$ whose name $t_{b}.n$ equals to the call name (e.g., \texttt{b}) of $call^\prime$ according to $\mathcal{C}.Incl$. If yes, the type of the callee function $f^\prime$ is inferenced; otherwise, it continues this procedure on the~super class of $t_{cls}$ according to $\mathcal{C}.Hier$. In other words, the ancestor classes of $t_{cls}$ are searched along the inheritance hierarchy. If $f^\prime$ is still not resolved, it searches $\mathcal{FTG}^f_{call^{\prime}.p}$ for the module type $t_m$ that is pointed to by the invoking variable (e.g., \texttt{a}) of $call^\prime$.~Then, it searches $\mathcal{F}$ for the function type $t_b$ that satisfies $t_b.n =\texttt{b}$ and $t_b.\phi = t_m.\phi.(t_m.n, \textbf{mod})$ (i.e., the function type~is~imported through module import expression \texttt{import...}). If found, the type of the callee function $f^\prime$~is~inferenced.




For the form of \texttt{b(...)}, it obtains its call name (e.g., \texttt{b}). Then, it searches $\mathcal{F}$ for the function type $t_b$ that satisfies $t_b.n = \texttt{b}$ and $t_b.\phi = f.\phi$ (i.e., the function type is in the same module with $f$). If such a $t_b$ is found, the callee function $f^\prime$ is resolved; otherwise, it continues to search $\mathcal{I}$ for the type relation $\langle t_s, t_t, e\rangle$~that~satisfies $t_s.n=\texttt{b}$ (i.e., the function type is imported by function import expression \texttt{from...import...}). If found, $t_t$ is the callee function's type, and $f^\prime$ is resolved.


\subsubsection{\textbf{Compute $\theta_{call^{\prime}}$}}\label{sec:approach:inter:theta} 

This step selects relevant type relations $\theta_{call^{\prime}}$ from the caller's FTG, and passes $\theta_{call^{\prime}}$ to the construction of the callee's FTG (Line 3 in Alg.~\ref{alg2}). In other words, $\theta_{call^{\prime}}$ is passed from the caller's inter-procedural analysis into the callee's intra-procedural analysis (Line 4). Specifically, $\theta_{call^{\prime}}$ contains type relations about invoking variable and argument variables of $call^\prime$, which can be directly selected from $\mathcal{FTG}^f_{call^{\prime}.p}$.

\subsubsection{\textbf{Compute $\Delta_{call^{\prime}}$}}\label{sec:approach:inter:delta} 

$\Delta_{call^\prime}$ is computed as the result of our inter-procedural analysis, which is passed back to the previous intra-procedural analysis (Line 5 in Alg.~\ref{alg2}). Given inputs as~$\mathcal{FTG}^{f^{\prime}}_{in}$~which is the FTG before intra-procedural analysis and $\mathcal{FTG}^{f^{\prime}}_{out}$ which is the FTG after intra-procedural analysis, this step computes changed type relations from $\mathcal{FTG}^{f^{\prime}}_{in}$ to $\mathcal{FTG}^{f^{\prime}}_{out}$, and puts them into $\Delta_{call^{\prime}}$. In particular, for each type relation $t_1 \xrightarrow{e_1} t_2$ in $\mathcal{FTG}^{f^{\prime}}_{out}$, if $t_1$ exists in a type relation $t_1 \xrightarrow{e_2} t_3$ in $\mathcal{FTG}^{f^{\prime}}_{in}$, it searches $\mathcal{FTG}^{f^{\prime}}_{out}$ for the final pointed type $t_n$ and adds a new type relation $t_1 \xrightarrow{e_3} t_n$ to $\Delta_{call^{\prime}}$; and similarly, if a type relation about the field of $t_1$ exists in $\mathcal{FTG}^{f^{\prime}}_{out}$ but does not exist $\mathcal{FTG}^{f^{\prime}}_{in}$, this type relation is also added to $\Delta_{call^{\prime}}$.

\begin{exmp}  
    \addnew{When evaluating the call to \texttt{redo()} of \texttt{Base.collector} in Fig.~\ref{fig:example_code}(a) at Line 11, there exists one type relation \texttt{Base.collector} $\xrightarrow{bpytop:10}$ \texttt{Cpu} selected from $\mathcal{FTG}^{process()}_{in}$. Therefore, the class type is inferenced as \texttt{Cpu}. Then, \texttt{redo} is resolved to a method in the class of type \texttt{Cpu}, resulting calling relation from \texttt{process()} to \texttt{Cpu.redo()} in Fig. \ref{fig:example_code}(c).}
\end{exmp}

%% file: src/ch04-evaluation-1.tex
\section{Evaluation}

To evaluate the efficiency and effectiveness of \tool, we design our evaluation to answer two research questions.

\begin{itemize}[leftmargin=*]
    \item \textbf{RQ1 Scalability Evaluation:} What is \tool's scalability compared with the state-of-the-art approach \textsc{PyCG}?
    \item \textbf{RQ2 Accuracy Evaluation:} What is \tool's accuracy compared with the state-of-the-art approach \textsc{PyCG}?
\end{itemize}

Further, we demonstrate a potential application of \tool~to~enhance dependency management through a case study.

\subsection{Evaluation Setup}

\todonew{\textbf{Benchmarks.} We use two benchmarks including a micro-benchmark (Table~\ref{table:micro_result}) and a macro-benchmark (Table~\ref{table:marcro_suite}). The micro-benchmark is extended from the dataset of \textsc{PyCG}~\cite{salis2021pycg} which contains 112~small Python programs covering a wide range of language features organized into 16 distinct categories. We extend it by respectively adding \todo{4}, \todo{4}, \todo{4}, \todo{5} and \todo{5} programs into category \texttt{context managers}, \texttt{arguments}, \texttt{assignments}, \texttt{direct calls}~and~\texttt{imports} to have more comprehensive coverage of the Python language features. Besides, to make our micro-benchmark support evaluation on flow-sensitivity, we create \todo{5} programs into a new category \texttt{control flow}. The new~categories are superscripted with ``$*$'' in Table~\ref{table:micro_result}.}

\todonew{For our macro-benchmark, we select \todo{6} Python applications under four criteria. For popularity, we query GitHub for popular Python projects that have over 1000 stars. For quality, we select projects that at least have development activities~in~the past year, and we filter out demo projects, cookbooks, etc.~For feasibility, we select real command-line projects that contain runnable test cases to facilitate the manual CGC. 
For scalability evaluation, we obtain the dependent libraries that are regarded as part of the whole program, and select projects whose whole program is more than 100k LOC and 10k functions. Table~\ref{table:marcro_suite} reports the statistics about the selected Python applications, where \textit{LOC (A.)} denotes the LOC in the application, \textit{LOC (W.)} denotes the LOC in the whole program, \textit{Func. (W.)} denotes the number of functions in the whole program, and \textit{Lib.} denotes the number of dependent libraries.}


\begin{table}[!t]
    \centering
    \footnotesize
    \caption{Statistics about Macro-Benchmark}\label{table:marcro_suite}
    \vspace{-5pt}
    \begin{tabular}{m{0.12cm}m{1.25cm}m{0.3cm}m{0.35cm}m{0.5cm}m{0.4cm}m{0.2cm}<{\centering}m{2.4cm}}   
    \noalign{\hrule height 1pt}
    Id. & Project & Stars  & LOC (A.)  & LOC (W.) & Func. (W.) & Lib. & Domain \\
    \noalign{\hrule height 1pt}
    P1 &bpytop & 9.0k & 5.0k & 120.5k   & 11.7k & 201   & resource monitor\\
    P2 & sqlparse & 3.1k  & 4.8k & 108.4k  & 10.8k & 191   & SQL parser module \\
    P3 & TextRank4ZH & 2.8k & 0.5k & 515.3k  & 30.0k & 237   &  text processing \\
    P4 &furl & 2.4k & 2.9k & 109.3k  & 11.0k & 190   & url manipulation \\
    P5 &rich-cli & 2.4k & 1.1k & 200.4k  & 18.6k & 192   & command~line~tool\\
    P6 & sshtunnel & 1.0k & 2.5k & 141.4k   & 14.5k & 202  & remote SSH tunnels \\
    \noalign{\hrule height 1pt}
    \end{tabular}
\end{table}

\textbf{RQ Setup.} The experiments are conducted on a MacBook Air with Apple M1 Chip and 8GB memory. \todel{We compare scalability and accuracy of \tool with \textsc{PyCG}.}\todel{As the result varies with regard to the scope of the analysis,}\todonew{We compare \tool with \textsc{PyCG} in \textbf{RQ1} and \textbf{RQ2} based on four scenarios}, \addnew{i.e., Exhaustive Application-Program (\textbf{E.A.}), Exhaustive Whole-Program (\textbf{E.W.}), Application-Centered Application-Program (\textbf{A.A.}), and Application-Centered Whole-Program (\textbf{A.W.}) CGC}. The functions in application-centered CGC in our experiment are a collection of application functions including those invoked from test cases, and those used in the Python module which would be automatically executed as a \texttt{main} when the module is being executed~or~imported.

\textbf{RQ1} is evaluated on our macro-benchmark. We compare the time and memory performance of \tool and \textsc{PyCG} using \textsc{UNIX} \textit{time} and \textit{pmap} commands. We conduct~this comparison only in E.A. and E.W. because \textsc{PyCG} intrinsically does not support \todonew{A.A. and A.W..}~Notice that \todel{\textsc{PyCG} also does not support whole-program CGC, and }we adapt~\textsc{PyCG} to support E.W.\todel{ (which is straightforward)} \textsc{PyCG} is analyzed in three configurations in E.A. and~E.W., i.e., running one iteration (e.g., E.W. (1)), running two iterations~(e.g., E.W. (2)), and waiting until it reaches~a~fixed-point (e.g., E.W. (m)). Besides, we adapt \tool to support E.A. and E.W. by taking all the functions in the analysis scope as entry functions. 

\todonew{Moreover, we assess the size of $\mathcal{FAG}$ in \tool and the assignment graph (AG) in \textsc{PyCG}, i.e., the number of points-to relations, to gain an internal and fine-grained understanding of the scalability results. $\mathcal{FAG}$ and AG represent the two intermediate components, respectively. For \tool, we compute the size of $\mathcal{FAG}$ in A.W., concerning the number of analyzed functions. For \textsc{PyCG}, we compute the size of AG in E.W., concerning the number of iterations. Additionally, we measure the proportion of the AG that is changed (i.e., added or deleted) after each~new~iteration.}

\textbf{RQ2} is evaluated on micro-benchmark and macro-benchmark. For micro-benchmark, we compare \tool and \textsc{PyCG} with the number of programs whose generated call graph is complete (C.) and sound (S.)\todel{ (C. for completeness and S. for soundness)}, and the correct, incorrect and missing call relations (TP, FP, and FN) in~different feature categories. A call graph is complete when it does not contain any call relation that is FP, and is sound when it does not have any call relation that is FN. For macro-benchmark, we compare the precision and recall of \tool and \textsc{PyCG} in E.A., A.A., and A.W., but we do not compare them in E.W. because of the potentially huge effort in constructing the ground truth. \todonew{Precision and recall are calculated using the ground truth call graph (i.e., $\mathcal{CG}_{GT}.E$) and the generated call graph (i.e., $\mathcal{CG}_{Gen}.E$), as formulated in Eq.~\ref{eq:accuracy}.} \todel{Precision is calculated by the proportion of correctly generated call relations (i.e., $\mathcal{CG}_{Gen}.E$ $\cap$ $\mathcal{CG}_{GT}.E$) in the generated call relations (i.e., $\mathcal{CG}_{Gen}.E$), and recall is calculated by the proportion of correctly generated call relations in the ground truth (i.e., $\mathcal{CG}_{GT}.E$), as formulated in Eq.~\ref{eq:accuracy}.}
\begin{equation}\label{eq:accuracy}
\begin{small}
\begin{aligned}
    \vspace{-5pt}
    Pre. = \frac{\mid  \mathcal{CG}_{Gen}.E \cap \mathcal{CG}_{GT}.E \mid}{\mid \mathcal{CG}_{Gen}.E \mid }
       \mathrm{,~} 
    Rec. =  \frac{\mid  \mathcal{CG}_{Gen}.E \cap \mathcal{CG}_{GT}.E \mid }{\mid \mathcal{CG}_{GT}.E \mid}
\end{aligned}
\end{small}
\end{equation}
As \textsc{PyCG} does not support A.A. and A.W., we use \textsc{PyCG}~to~generate call graphs in E.A. and E.W., and prune the call graphs by only keeping the functions that are reachable from the entry functions.

\begin{table*}[!t]
    \centering
    \footnotesize
    \caption{Scalability Results on Our Macro-Benchmark (\textit{T.} denotes time in seconds, \textit{M.} denotes memory in MB if not specified)}\label{table:marcro_scalability}
    \vspace{-5pt}
    \begin{tabular}{m{0.2cm}m{0.5cm}m{0.5cm}m{0.5cm}m{0.5cm}m{0.5cm}m{0.5cm}m{0.7cm}m{0.6cm}m{0.5cm}m{0.5cm}m{0.5cm}m{0.5cm}m{0.6cm}m{0.6cm}m{0.5cm}m{0.5cm}m{0.5cm}m{0.5cm}}   
    \noalign{\hrule height 1pt}
    \multirow{3}{*}{Id.} & \multicolumn{10}{c}{\textsc{PyCG}} & \multicolumn{8}{c}{\tool} \\
    & \multicolumn{2}{c}{E.A. (1)}& \multicolumn{2}{c}{E.A. (m)} &  \multicolumn{2}{c}{E.W. (1)} &  \multicolumn{2}{c}{E.W. (2)} &  \multicolumn{2}{c}{E.W. (m)} & \multicolumn{2}{c}{E.A.} & \multicolumn{2}{c}{E.W.}  & \multicolumn{2}{c}{A.A.} & \multicolumn{2}{c}{A.W.} \\
    \cmidrule(lr){2-11}
    \cmidrule(lr){12-19}
    & T. & M. & T. & M.& T. & M.& T. & M. & T. & M. & T. & M.  & T. & M. & T. & M. & T.& M. \\
    \noalign{\hrule height 1pt}
    P1 &0.78&79&1.26&90&78.49&766&113.50&799&\cellcolor{lightgray} 24h+&2.3G&1.22&61&48.43&1003&0.88&53&\todonew{7.23}& \todonew{190}
    \\
    P2 &0.55&36&1.01&40&48.83&702&76.29&855&\cellcolor{lightgray} 24h+&5.7G&0.46&33&33.79&764&0.31&30&\todonew{2.69}&\todonew{82}
    \\
    P3&0.16&24&0.19&24&1705.07&1630&1947.39&\cellcolor{lightgray} OOM&4h+&\cellcolor{lightgray} RE&0.19&24&1465.26&2061&0.18&24&\todonew{19.14}&\todonew{407}
    \\
    P4 &0.25&30&0.35&30&47.14&569&72.53&636&\cellcolor{lightgray} 24h+&5.1G&0.30&32&34.35&775&0.22&29&\todonew{1.92}&\todonew{65}
    \\
    P5 &0.20&26&0.25&28&988.36&1431&2190.40&\cellcolor{lightgray} OOM&0.5h+&\cellcolor{lightgray} OOM&0.26&26&163.01&1523&0.21&25&\todonew{12.64}&\todonew{408}
    \\
    P6 &0.19&28&0.24&29&149.02&684&245.13&741& \cellcolor{lightgray} 24h+&4.6G&0.25&26&62.69&1250&0.19&26&\todonew{5.38}&\todonew{213}
    \\\hline
    Avg. &0.36&37&0.55&40&502.82&963&774.21&757&16.7h+&4.4G+&0.45&33&301.26 &1229&0.33&31&\todonew{8.16}&\todonew{227}
    \\
    \noalign{\hrule height 1pt}
    \end{tabular}
\end{table*}

\textbf{Ground Truth Construction.} \todonew{For micro-benchmark, we reuse the ground truth from \textsc{PyCG}~\cite{salis2021pycg} and we build the ground truth manually for our newly-added programs in micro-benchmark.} For macro-benchmark, we first execute the test cases and collect call traces for each~application \todonew{by executing }\todel{using the embedded Python trace module }\textit{``python -m trace --listfuncs $<$python\_file$>$''}.\todel{ The call traces span across the whole program. } Then, the call traces are transformed into the same format in the micro-benchmark. The transformed call graph contains implicit call relations that are invisible and inherently invoked by the Python interpreter; e.g., with \texttt{import} keyword, the Python interpreter invokes functions from \texttt{\_fronzen\_importlib} which is not part of the whole program. We filter them out from the ground truth. Second, we enlarge the collection of call relations generated by test case execution by manually inspecting application functions (i.e., functions in application modules exclusive of library functions). Specifically, we go through the code of each application function and add missing call relations. This step is to improve the \todonew{incomplete coverage of CGC}\todel{imperfect construction result of call graph} using test cases, \todonew{e.g., the missing branches while executing test cases.}\todel{complete test cases might miss call relations.} Three of the authors are involved in this procedure, which takes~\todo{6}~person~months.

In summary, we construct the ground truth of call graphs with a total number of \todo{5,653} functions and \todo{20,085} call relations. 


%% file: src/ch04-evaluation-2.tex

\subsection{Scalability Evaluation (RQ1)}



The scalability results in terms of time and memory performance on our macro-benchmark are reported in Table~\ref{table:marcro_scalability}. 

In terms of time, \tool and \textsc{PyCG} generate a call graph in E.A. within a second. \textsc{PyCG} takes more time for more iterations in E.A. (i.e., from \todo{0.36} seconds in E.A. (1) to \todo{0.55} seconds in E.A. (m)). The gap becomes larger in E.W., where \textsc{PyCG} takes averagely~\todo{502.82}~seconds in E.W. (1), and \tool takes averagely \todo{301.26}~seconds. As the iteration increases, \textsc{PyCG} crashes in two projects due to out-of-memory (OOM) and recursion error (RE). Specifically, \textsc{PyCG} runs out-of-memory in E.W. (2) and suffers recursion error in E.W. (m) on P3, while \textsc{PyCG} runs out-of-memory in E.W. (2) and E.W. (m) on P5. We record the consumed time immediately before \textsc{PyCG} is crashed. Thereby, the average time for \textsc{PyCG} in E.W. (2) and E.W. (m) is more than \todo{774.21} seconds and more than \todo{16.7} hours. Therefore, \tool runs \todo{67\%} faster than \textsc{PyCG} in E.W. (1), and at least \todo{157\%} faster than \textsc{PyCG} in E.W. (2). Furthermore, we also measure time for \tool's application-centered CGC. \tool takes \todo{0.33} seconds for A.A. and \todonew{8.16} seconds for A.W., which significantly consumes less time than in the exhaustive generation.


In terms of memory, \tool consumes \todo{4 MB} less memory than \textsc{PyCG} in E.A. (1), and \todo{7 MB} less memory than \textsc{PyCG} in E.A. (m). When the analysis scope is expanded to the whole program, \textsc{PyCG}~consumes \todo{266 MB} less memory than \tool in E.W. (1). However, when the iteration number increases, \textsc{PyCG} consumes significantly more memory than \tool, and also suffers out-of-memory and recursion error in E.W. (2) and E.W. (m). Moreover, it takes \tool \todo{31 MB} in A.A. and \todonew{227 MB} in A.W., which significantly consumes less memory than in exhaustive CGC.



\begin{figure}[!t]
    \centering
    \begin{subfigure}[b]{0.22\textwidth}
        \centering
        \includegraphics[scale=0.25]{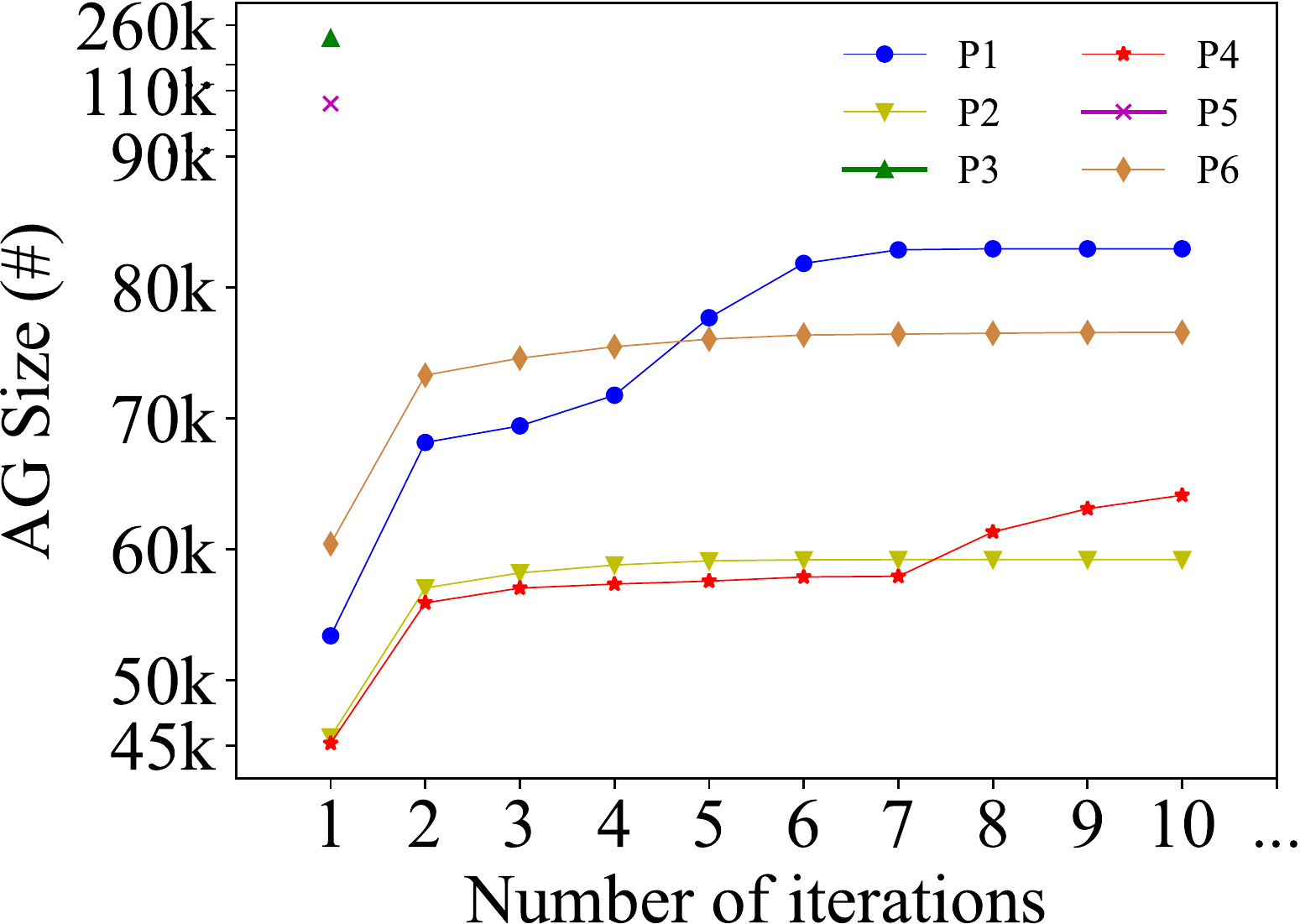}
        \vspace{-13pt}
        \caption{\todonew{AG Size}}
    \end{subfigure}
    \begin{subfigure}[b]{0.22\textwidth}
        \centering
        \includegraphics[scale=0.256]{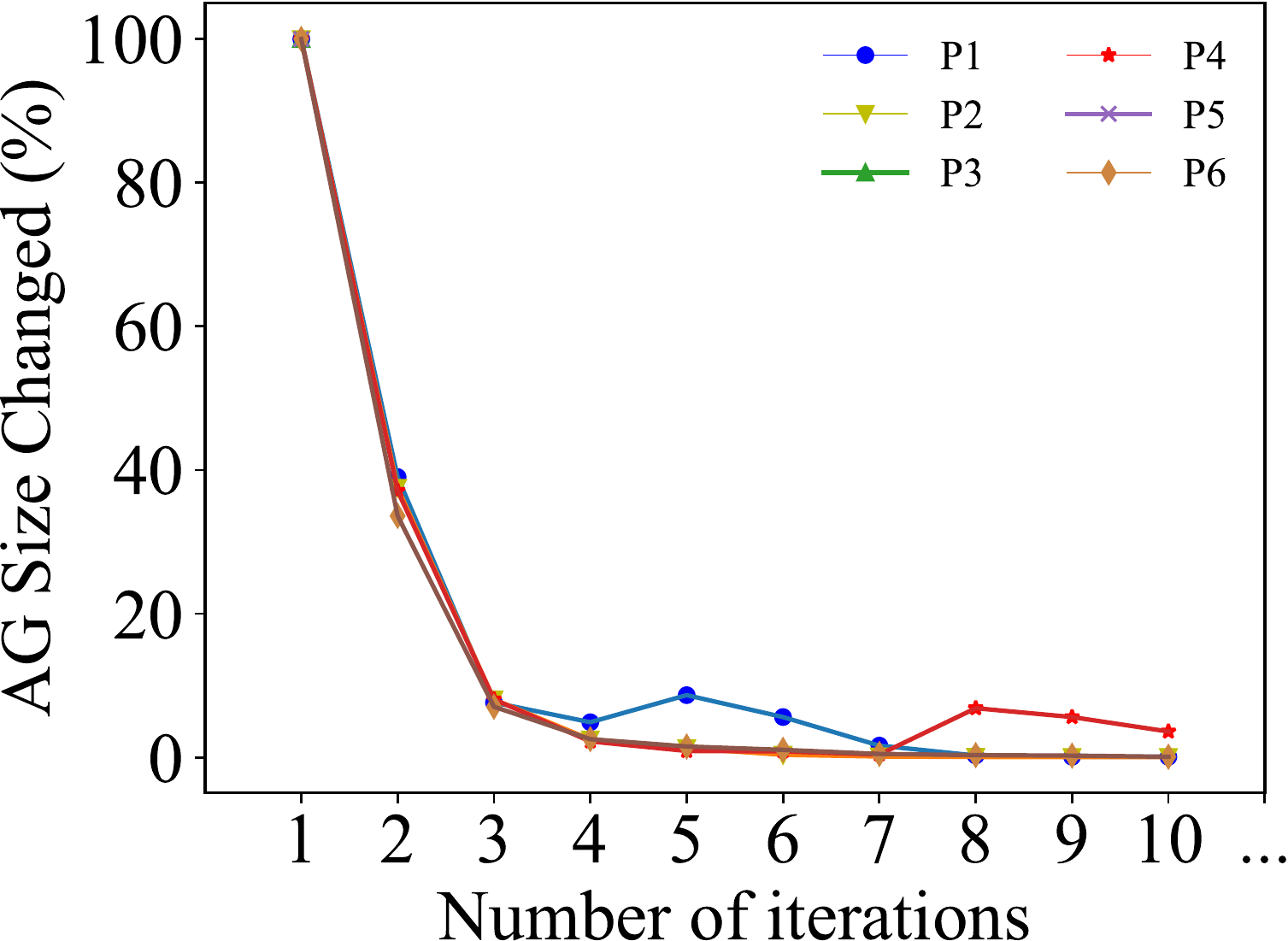}
        \vspace{-13pt}
        \caption{\todonew{AG Size Changed (\%)}}
    \end{subfigure}
    \vspace{-3pt}
    \caption{The Size and Changed Proportion of the Assignment Graph (AG) in \textsc{PyCG} w.r.t the~Number~of~Iterations}\label{fig:agsize}
\end{figure}

\begin{figure}
    \centering
    \vspace{-2.5pt}
    \includegraphics[scale=0.5]{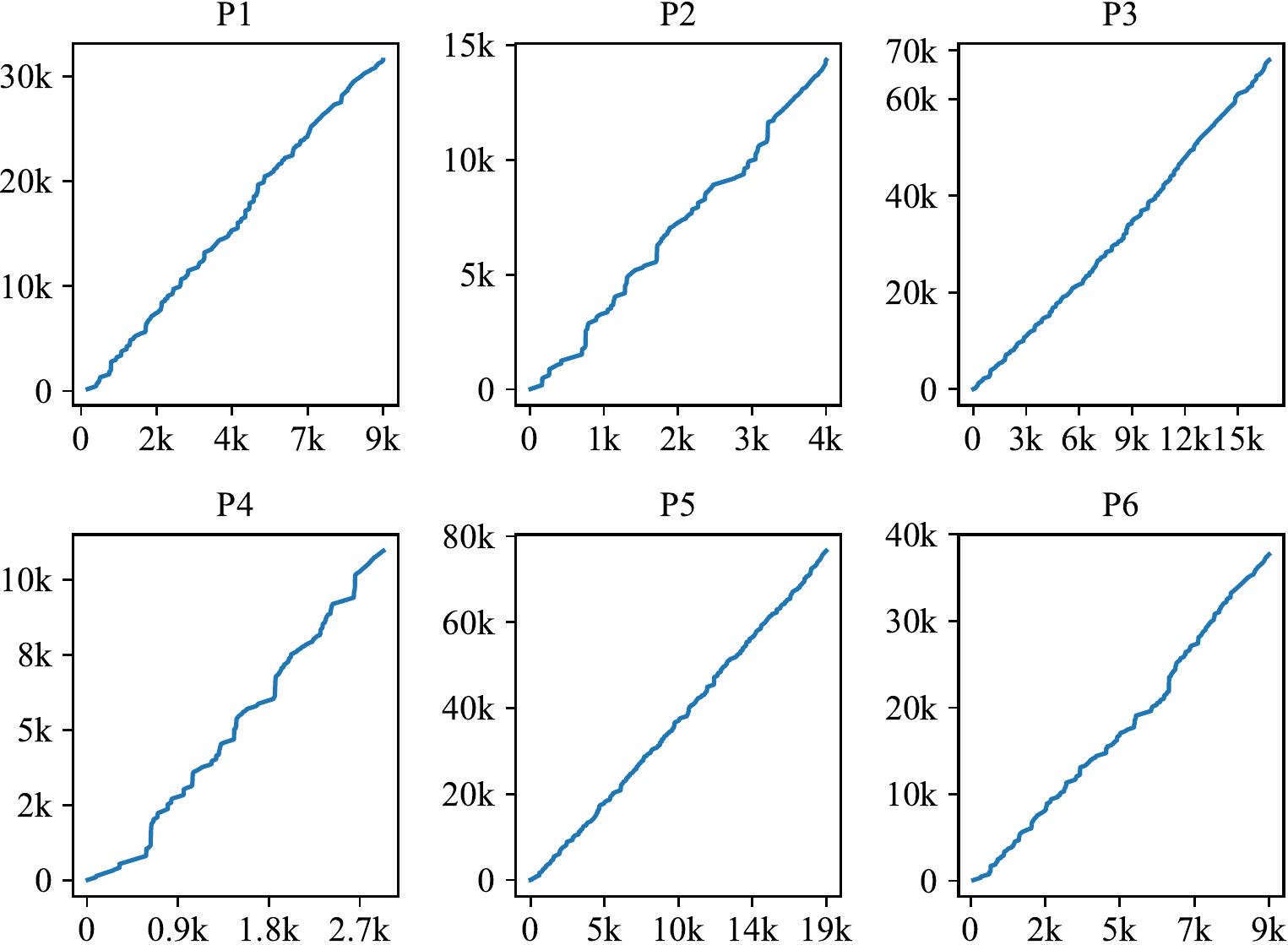}
    \vspace{-3pt}
    \caption{\todonew{The Total Size of the $\mathcal{FTG}$s in \tool w.r.t the Number of Analyzed Functions}}
    \label{fig:fagsize} 
\end{figure}

\todonew{Further, we report the size of AG in \textsc{PyCG} in Fig.~\ref{fig:agsize} and~the~size of $\mathcal{FTG}$ in \tool in Fig.~\ref{fig:fagsize}. Specifically, Fig.~\ref{fig:agsize}(a) illustrates that \textsc{PyCG} generates a substantial number of points-to relations during the first iteration, ranging from \todo{45k} to \todo{257k}. Subsequent iterations show a continuous increase in the number of relations, with marginal increases after the 10th iteration. Notably, for P3 and P5, there is only a single dot representing the initial AG size before OOM occurs. For the other four projects, the AG size ranges from \todo{59k} to \todo{82k} after the 10th iteration. Additionally, Fig.~\ref{fig:agsize}(b) reveals that the proportion of changed AG decreases significantly as the iterations progress. After the 3rd iteration, the proportion of changed AG drops to a value of less than \todo{10\%}, suggesting that a majority of the computation may be unnecessary. In contrast, Fig.~\ref{fig:fagsize} demonstrates that the size of $\mathcal{FTG}$ generated by \tool scales proportionally with the number of analyzed functions as more functions are processed. On average, \tool analyzes \todo{10,124} functions and maintains $\mathcal{FTG}$s with a total size of \todo{39,714}. In comparison to the average AG size of \todo{107,499} in \textsc{PyCG}, the total size is \todo{63\%} smaller. This observation suggests that employing application-centered CGC could significantly reduce both the computational scale and memory requirement.}

\begin{tcolorbox}[size=title, opacityfill=0.15]
\textit{\textbf{Summary.}} \tool runs \todo{67\%} faster than \textsc{PyCG} in E.W. (1) and at least \todo{157\%} faster than \textsc{PyCG} in E.W. (2). It only takes \tool averagely \todonew{8.16} seconds to generate application-centered whole-program call graphs (i.e., in A.W.). \tool is memory-efficient in both E.W. and A.W. Therefore, \tool is scalable.
\end{tcolorbox}

%% file: src/ch04-evaluation-3.tex

\subsection{Accuracy Evaluation (RQ2)}

We present the accuracy results of \tool and \textsc{PyCG} on our micro-benchmark and macro-benchmark.

\textbf{Micro-Benchmark.} The accuracy results on our micro-benchmark are presented in Table~\ref{table:micro_result}. In terms of completeness, \textsc{PyCG} generates call graphs that are complete for \todo{113} programs. \todo{20} of the incomplete cases come from our newly-added categories (superscripted with ``$*$'') and the rest \todo{5} incomplete cases come from the original category. Those incomplete cases generate call graphs with \todo{27} false positives (FP). The majority (\todo{21}) of the \todo{27} false positives for \textsc{PyCG} are located in the five newly added categories. Differently, \tool generates complete call graphs for \todo{132} cases. \todel{\tool only generates \todo{one} incomplete case with \todo{one} false positive in \texttt{decorators}.} \todonew{The incomplete cases of \tool span in categories such as \texttt{dicts} and \texttt{lists}.}
In terms of soundness, \textsc{PyCG} generates call graphs that are sound for \todo{126} programs. The rest \todo{12} unsound cases span in categories~such as \texttt{context-managers}, \texttt{built-ins}, \texttt{control flow}, \texttt{assignment}, \texttt{dicts} and \texttt{lists}. The unsound cases cause \todo{27} false negatives. There are \todo{7} false negatives in \texttt{built-ins}, ranking at the top among other categories. In the meantime, \tool generates \todo{134} sound call graphs. The unsound cases of \tool span in categories such as \texttt{assignments}, \texttt{built-ins},  \texttt{dicts} and \texttt{lists}. \todel{\tool generates \todo{35} false negatives, while the majority (\todo{26}) of the false negatives come from \texttt{dicts} and \texttt{lists}.}

\textbf{Macro-Benchmark.} The accuracy results on our macro-benchmark are presented in Table~\ref{table:marcro_accuracy}. 
In terms of precision, \tool achieves~similar precision in E.A. and A.A. compared with \textsc{PyCG}. Besides, \textsc{PyCG} achieves similar precision in E.A. (1) and E.A. (m), and A.A. (1) and A.A. (m) due to the relatively small analysis scope.~In~A.W., \textsc{PyCG}'s precision drastically drops to \todo{0.19}, while \tool's precision also greatly drops to \todo{0.35}, but is \todo{84\%} higher than \textsc{PyCG}. 
\todonew{The reason for the drop is that} in exhaustive analysis, accuracy is computed by comparing sets of call relations, whereas in application-centered CGC, accuracy is computed by comparing chains of call relations from entry functions.~In other words, if there exists one false positive call relation, the subsequent call relations along the generated call chain are all considered as false positives. Therefore, for application-centered CGC, the impreciseness of the call graph is magnified. \todonew{Nevertheless, \tool achieves a higher precision than \textsc{PyCG}.}


Moreover, we inspect the precision loss of \textsc{PyCG} against \tool. The major reason is that \textsc{PyCG} reports call relations disregarding control flows, whereas \tool is flow-sensitive. In addition, \text{PyCG} also reports false positives due to insufficient treatment of class inheritance. 
\todel{For example, for function calls invoked from base classes, it reports function calls in sub classes, but they do not exist in the sub classes.} \todonew{For example, \textsc{PyCG} reports non-existing function calls in sub classes instead of in base classes.} \todel{We inspect the impreciseness of \tool.} The major reason \todonew{of \tool's impreciseness} is correlated with \todonew{unresolved function definitions.} If the points-to relations are not updated timely because of the unresolved function definitions, it would report false call relations because of the outdated~points-to~relations.



\begin{table}[!t]
    \small
    \centering
    \footnotesize
    \caption{Accuracy Results on Our Micro-Benchmark}\label{table:micro_result}
    \begin{tabular}{m{1.3cm}m{0.6cm}m{0.6cm}m{0.1cm}m{0.1cm}m{0.1cm}m{0.6cm}m{0.6cm}m{0.1cm}m{0.1cm}m{0.1cm}}   
    \noalign{\hrule height 1pt}
    \multirow{2}{*}{Category} &  \multicolumn{5}{c}{\textsc{PyCG}} & \multicolumn{5}{c}{\tool } \\
    \cmidrule(lr){2-6}
    \cmidrule(lr){7-11}
    & C. & S. & TP & FP & FN  &C. &S. &  TP & FP & FN \\ 
    \noalign{\hrule height 1pt}
    arguments & 6/6 & 6/6&14& 0& 0  & 6/6& 6/6&14& 0& 0 \\
    assignments & 4/4& 3/4&13& 0& 2  & 4/4& 3/4&13& 0& 2 \\
    built-ins & 3/3& 1/3&3& 0& 7  & 3/3& \textbf{2/3} & \textbf{6} & 0& \textbf{4} \\
    classes & 22/22& 21/22&65& 0& 1  & 22/22& 21/22&64& 0& 2 \\
    decorators & 6/7& 7/7&22& 1 & 0 & \textbf{7/7} & 7/7&22& \textbf{0}& 0\\
    dicts & \textbf{12/12} & 11/12&21& 0& 2 & 11/12& \todonew{12}/12& \textbf{\todonew{23}} & \textbf{\todonew{1}}& \textbf{\todonew{0}} \\ 
    direct calls & 4/4& 4/4&10& 0& 0  & 4/4& 4/4&10& 0& 0 \\
    exceptions & 3/3&3/3& 3& 0& 0 & 3/3&3/3& 3& 0& 0 \\
    functions & 4/4& 4/4&4& 0& 0  & 4/4& 4/4&4& 0& 0\\
    generators &6/6& 6/6&\textbf{18}& 0& 0  & \todonew{5/6}& \todonew{6/6}& 18& 1& 0 \\
    imports & 13/14& 14/14&20& 2 & 0 & \textbf{14/14}& 14/14&20& \textbf{0}& 0 \\
    kwargs & 2/3& 3/3&9& 1& 0  & \textbf{3/3} & 3/3&9& \textbf{0} & 0  \\
    lambdas & 5/5& 5/5&14& 0& 0  & 5/5& 5/5&14& 0& 0 \\
    lists & \textbf{7/8} & 6/8& 14& \textbf{1}& 2  & 4/8& \textbf{7/8} &\textbf{15}& 5& \textbf{1} \\
    mro & 6/7& 6/7&19& 1& 1 & \textbf{7/7} & \textbf{7/7} &\textbf{20} & \textbf{0} & \textbf{0} \\
    returns & 4/4& 4/4& 12& 0& 0 & 4/4& 4/4&12& 0& 0\\
    \addnew{context managers$^*$} & 4/4 & 0/4 & 3 & 0 & 12 & 4/4& 4/4& 15& 0 & 0 \\
    arguments$^*$ & 0/4& 4/4&8& 4& 0 & \textbf{4/4} & 4/4&8& \textbf{0}& 0 \\
    assignments$^*$ & 0/4& 4/4&4& 4& 0& \textbf{4/4} & 4/4&4& \textbf{0}& 0 \\
    direct calls$^*$ & 0/5& 5/5&18& 5& 0 & \textbf{5/5}& 5/5&18& \textbf{0}& 0\\
    imports$^*$ & 1/5& 5/5&7& 4& 0 & \textbf{5/5}& 5/5&7& \textbf{0}& 0\\
    control flow$^*$ & 1/4& 4/4&16& 4& 0  & \textbf{4/4} & 4/4&\textbf{16}& \textbf{0} & 0 \\\hline
    Total & \todonew{113/138}  & 126/138  & 317& 27& 27  & \textbf{132/138}  & \textbf{134/138} & \textbf{\todonew{335}} & \textbf{7}& \textbf{9} \\
    \noalign{\hrule height 1pt}
    \end{tabular}
\end{table}

%



In terms of recall, \tool achieves a recall of \todo{0.82} and \todo{0.66} in E.A. and A.A., improving \textsc{PyCG} by \todo{8\%} in both E.A. and A.A. In A.W. (1), \tool improves \textsc{PyCG} by \todo{15\%}, while in A.W. (2), \tool improves \textsc{PyCG} by \todo{10\%} on the four projects where \textsc{PyCG} does not crash. The recall of \textsc{PyCG} and \tool drops from E.A. to A.A., which is also mainly caused by the accuracy computing difference. \todo{The same reason also explains the recall drop from A.A. to A.W..}\todel{; i.e., in application-centered CGC, the larger the analysis scope, the longer the call chains, and the higher chance to introduce more false negatives.}

Furthermore, we inspect the recall gain of \tool against \textsc{PyCG}. The major reason is that \tool supports \texttt{built-ins} and \texttt{control flow} more comprehensively than \textsc{PyCG}. For example, regarding support for \texttt{built-ins}, \textsc{PyCG} misses the call \texttt{builtins.split} for the call expression \textit{`tel-num'.split(`-')}; and regarding \texttt{control flow}, \textsc{PyCG} does not handle the \texttt{with} statements, and thus ignores the relevant control flows. We also inspect the false negatives of \tool, which are mainly caused by two reasons. First, functions stored in \texttt{list}, \texttt{tuple} and \texttt{dict} are missing as our FAG does not maintain points-to relations for the above data structures. Second, functions invoked from dynamic linked libraries (e.g., \textit{math-python-*.so}) are missing due to reflective invocations and \tool's incapability of inter-procedural analysis of dynamic linked libraries.





\begin{table*}[!t]
    \small
    \centering
    \footnotesize
    \caption{Accuracy Results on Our Macro-Benchmark}\label{table:marcro_accuracy}
    \begin{tabular}{m{0.4cm}m{0.53cm}m{0.53cm}m{0.53cm}m{0.53cm}m{0.53cm}m{0.53cm}m{0.53cm}m{0.53cm}m{0.53cm}m{0.53cm}m{0.53cm}m{0.53cm}m{0.53cm}m{0.53cm}m{0.53cm}m{0.53cm}m{0.53cm}m{0.53cm}}   
    \noalign{\hrule height 1pt}
    \multirow{3}{*}{Id.} & \multicolumn{12}{c}{\textsc{PyCG}} & \multicolumn{6}{c}{\tool} \\
    & \multicolumn{2}{c}{E.A. (1)} & \multicolumn{2}{c}{A.A. (1)} & \multicolumn{2}{c}{E.A. (m)} & \multicolumn{2}{c}{A.A. (m)} & \multicolumn{2}{c}{A.W. (1)}  & \multicolumn{2}{c}{A.W. (2)}  & \multicolumn{2}{c}{E.A.} & \multicolumn{2}{c}{A.A.} & \multicolumn{2}{c}{A.W.} \\
    \cmidrule(lr){2-13}
    \cmidrule(lr){14-19}
    & Pre. & Rec.  & Pre. & Rec.  & Pre. & Rec. & Pre. & Rec. & Pre. & Rec. & Pre. & Rec. & Pre. & Rec. & Pre. & Rec. & Pre. & Rec. \\
    \noalign{\hrule height 1pt}
    P1 &0.99&0.84&1.00&0.83&0.99&0.84&1.00&0.83&0.20&0.60&0.20&0.60&0.99&0.92&1.00&0.93&0.27&0.66
    \\
    P2 &0.99&0.62&1.00&0.32&0.99&0.62&1.00&0.32&0.19&0.48&0.19&0.49&0.99&0.64&1.00&0.35&0.42&0.64
    \\
    P3 &1.00&0.87&1.00&0.90&1.00&0.87&1.00&0.90&0.20&0.28&-&-&1.00&0.95&1.00&0.90&0.29&0.34
    \\
    P4 &0.99&0.60&1.00&0.31&0.99&0.60&1.00&0.31&0.15&0.49&0.15&0.49&1.00&0.63&1.00&0.31&0.49&0.61
    \\
    P5 &0.96&0.82&1.00&0.78&0.96&0.82&1.00&0.78&0.14&0.51&-&-&1.00&0.95&1.00&0.94&0.19&0.65
    \\
    P6 &1.00&0.80&1.00&0.52&1.00&0.80&1.00&0.52&0.23&0.51&0.23&0.52&1.00&0.85&1.00&0.56&0.34&0.68
    \\\hline
    Aver. &0.99&0.76&1.00&0.61&0.99&0.76&1.00&0.61&0.19&0.48&0.19&0.52&0.99&0.82&1.00&0.66&0.35&\todonew{0.60}
    \\
    \noalign{\hrule height 1pt}
    \end{tabular}
\end{table*}

\begin{tcolorbox}[size=title, opacityfill=0.15]
\textit{\textbf{Summary.}} \tool achieves similar precision to \textsc{PyCG} in E.A. and A.A., but improves \textsc{PyCG} in recall by \todo{8\%} in E.A. and~A.A. 
Further, \tool obtains a precision of \todo{0.35} and a recall~of~\todonew{0.60} in A.W., which significantly improves \textsc{PyCG} by \todo{84\%} in precision and at least \todonew{20\%} in recall.
\end{tcolorbox}

\subsection{Case Study: Fine-Grained Tracking of Vulnerable Dependencies}

Dependency management tools (e.g., Dependabot~\cite{dependabot} provided by GitHub) notify developers about vulnerable dependencies used~in~a project. However, these tools do not analyze whether the project~invokes a vulnerable function. Such a reachability analysis requires scalable whole-program CGC which is missing~in the literature. We show that \tool can be used to equip dependency management tools with vulnerability reachability analysis~so~that developers can give a high priority to fixing vulnerable dependencies whose vulnerable functions are reachable through call chains.

To this end, we conduct the following steps to demonstrate~the~usefulness of \tool in this context. First, we select \todo{43} Python projects from the top 200 highly-starred GitHub projects by removing demos, learning guides, paper collections, etc. These \todo{43} projects~directly or transitively depend on \todo{276} Python dependencies. Second, we choose the top \todo{10} most commonly used dependencies. These~dependencies are utilized in \todo{30} projects. We query GitHub advisory~\cite{githubadvisory}~to obtain vulnerabilities in these dependencies. As a result, we collect \todo{36} vulnerabilities in \todo{6} dependencies, affecting~\todo{9}~projects. 




 
Finally, we use \tool to generate call graphs for these \todo{9} affected projects in A.W. mode, and analyze whether the vulnerable functions (identified by vulnerability patches) are reachable through~call chains. 
\todonew{\tool successfully generates call graphs for all projects.}
It turns out that \todo{6} vulnerabilities in \todo{one}~dependency~are~reachable through \todo{6} call chains in \todo{6}~projects.~We~manually~confirm that the \todo{6} call chains generated by \tool~are~all~correct. \todel{The length~of the call chains ranges from a minimum of \todo{3}~to~a~maximum of \todo{7}.}

\todonew{\textbf{Comparison to \textsc{PyCG}.} We evaluate \textsc{PyCG} on the \todo{9} projects.~However, it suffers OOM on \todo{3} projects. 
\textsc{PyCG} finds that \todo{4} vulnerabilities in \todo{one} dependency are reachable through call chains in~\todo{4}~project.}



%


\begin{tcolorbox}[size=title, opacityfill=0.15]
\textit{\textbf{Summary.}}
Of the \todo{36} vulnerabilities in \todo{6} dependencies that are used in \todo{9} projects, \tool helps to identify \todonew{6} vulnerabilities in \todo{one} dependency that are reachable via \todonew{6} call chains in \todonew{6} projects. This fine-grained analysis guides developers to focus on reachable and thus more risky vulnerable dependencies.
\end{tcolorbox}

\subsection{Threats}

The primary threats to the validity of our experiments are our benchmark and the construction of ground truth. For micro-benchmark,~we add \todo{23} programs that are written by two authors that have at least two years of Python programming experience to have a more complete coverage of language features. Although it is exhaustive, we believe it is representative. Besides, the ground truth of these 25 programs is easy to build because they are~all~simple programs. For macro-benchmark, we carefully select 6 real-world Python applications. We believe they are representative because they are popular in the community, well-maintained, feasible to run, and large-scale in whole-program. However, the construction of their ground truth is a challenging task. Therefore, we construct it in two ways, i.e., automated test case execution, and manual investigation. The manual investigation is conducted by two of the authors independently, and another author is involved in resolving disagreements. We do not use the macro-benchmark of \textsc{PyCG} as it only contains call relations in application code.~To~the~best of our knowledge, our macro-benchmark is the largest one.

%% file: src/ch05-related-work.tex

\section{Related Work}




\textbf{Points-to Analysis.} Many research works have proposed to improve the precision~of~points-to analysis from different perspectives, e.g., combining call-site sensitivity with object sensitivity~\cite{kastrinis2013hybrid, lu2021eagle, lhotak2008evaluating}, demand-driven points-to analysis~\cite{sui2017demand, sridharan2005demand, spath2016boomerang, sui2017demand}, combining type-analysis with points-to analysis~\cite{allen2015combining}, selectively applying context-sensitive analysis w.r.t precision-critical methods~\cite{li2018precision, lu2021selective}. Hardekopf et al.~\cite{hardekopf2007ant} propose to collapse inclusion constraint cycles based on inclusion-based pointer analysis~\cite{andersen1994program} to reduce the scalability of online cycle detection. Wilson et al.~\cite{wilson1995efficient} summarize the effects of a procedure using partial transfer function (PTF)\todel{, thus enabling context-sensitive analysis without reanalyzing at every call site}. 
Our approach uses flow-sensitive points-to analysis by encoding point-to relations with flow information in the function assignment~graph.\todel{ and uses demand-driven analysis to skip the analysis of~unreachable~functions.}


\textbf{Type Inference for Dynamic programming languages.} Type inference is a classic topic in programming language research. For dynamic programming languages, the types of the variables are not determined until runtime execution, which renders type inference more demanded. PySonar2~\cite{Pysonar2} and Pyr~\cite{Pyre} are two popular unification-based type inference tools for Python. PyAnnotate~\cite{pyannotate} generates type annotations based on arguments and return types observed at runtime. Salib proposes Starkille~\cite{salib2004starkiller}, which includes an External Type Description Language that enables extension authors to document how their foreign code extensions interact with Python. Xu et al. propose Pyttrace~\cite{xu2016python}, which uses attribute accesses and variable names to infer variable types. Differently, \tool uses a static and direct type inference approach. Instead of collecting attributes access and variable names, \tool directly infers the type of variables based on the assignment statements using flow-sensitive analysis. While conducting the flow-sensitive analysis, \tool makes strong updates on variables to make type inference more precise.

\textbf{Call Graph Construction (CGC).} \todonew{CGC methods typically correlate with the characteristics of programming languages.} \todel{CGC methods cover a variety of programming languages, e.g., Java~\cite{vallee2010soot, watson2017libraries}, Python~\cite{salis2021pycg}, JavaScript~\cite{nielsen2021modular}, C/C++~\cite{milanova2004precise, wilson1995efficient, andersen1994program}, Scala~\cite{petrashko2016call}, etc.} In Java, \todel{to deal with the missing edges caused by their language features, }many approaches are proposed to handle language features such as reflection~\cite{sun2021taming, sawin2011assumption, bodden2011taming, liu2017reflection, sui2017use, grech2017heaps}, foreign function~interface~\cite{lee2020broadening},~dynamic~method invocation\cite{bodden2012invokedynamic},~serialization~\cite{santos2021serialization}, and library call~\cite{zhang2007automatic}. To deal with scalability issues, Ali et al. propose CGC~\cite{ali2012application} and Averroes~\cite{ali2013averroes} to approximate the call graph by considering application-only. \todel{CGC and Averroes generate a quickly-constructed placeholder library that over-approximates the possible behavior of an original library, therefore avoiding the big cost of whole-program analysis. }\todel{Such application-program analysis omits the inner calls from the library which may not be useful in some static analysis tasks (e.g., vulnerability propagation analysis).} In the face of \textit{application-centered} needs, some approaches consider only the \textit{influenced nodes} compared to those exhaustive ones for the entire program~\cite{agrawal2002evaluating, heintze2001demand, agrawal2000demand, spath2016boomerang}. Reif~\cite{reif2016call} proposes CGC for potentially exploitable vulnerabilities. In JavaScript, many approaches are proposed to handle reflection~\cite{schafer2013dynamic}, event-driven~\cite{madsen2015static}, and scalability issues\cite{feldthaus2013efficient}. \todel{To deal with scalability issues, Feldthaus et al.~\cite{feldthaus2013efficient} approximate call graph by using field-based analysis, tracking method objects, and ignoring dynamic property accesses which simplifies type analysis of variables, making the flow analysis lightweight.} Madsen et al.~\cite{madsen2013practical} combine pointer analysis with use analysis to save the effort of analyzing library API bodies. In Scala, Petrashko et al.~\cite{petrashko2016call} and Ali et al.~\cite{ali2015type} handle polymorphism to enhance~CGC.

In Python, Salis et al.~\cite{salis2021pycg} propose \textsc{PyCG} to generate a practical call graph that could efficiently and accurately generate call graph edges by handling several Python features. 
\textsc{PyCG} outperforms \textsc{Pyan}~\cite{pyan} and \textsc{Depends}~\cite{depends} in CGC. Besides, Hu et. al.~\cite{hu2021static} propose to link the missing edges~caused by foreign function~interfaces in Python. However, CGC for Python is not scalable to application with hundreds of dependent libraries. \todo{To overcome the drawbacks in Python CGC, we propose a \textit{application-centered} CGC approach~for~Python.}

\textbf{Evaluating Call Graph Construction.} Several studies have evaluated CGC tools for various languages such as C~\cite{murphy1998empirical}, Java~\cite{reif2018systematic, reif2019judge, ali2019study}, and JavaScript~\cite{antal2018static}. 
In Java ecosystem, Sui et al.~\cite{sui2018soundness} construct~a~micro-benchmark~for dynamic language features to evaluate the soundness~of Soot~\cite{vallee2010soot}, WALA~\cite{watson2017libraries} and Doop~\cite{bravenboer2009strictly}. Besides, Sui et al.~\cite{sui2020recall} also evaluate the soundness of Doop with real-world Java programs. Reif et al.~\cite{reif2018systematic, reif2019judge} construct an extensive test suite covering~language and API features in Java, and compare the soundness~and~time overhead of Soot, WALA, OPAL\cite{eichberg2018lattice}~ and Doop. Differently, Tip~and Palsberg~\cite{tip2000scalable} explore the precision and scalability of different approaches for resolving virtual function calls (e.g., CHA~\cite{dean1995optimization}, RTA~\cite{bacon1996fast} and $k$-CFA~\cite{allen1970Cfa, shivers1991control}). To the best of our knowledge,~none of these studies address the scalability problem in Python language, and our work fills this gap.

%% file: src/ch07-conclusion.tex
\section{Conclusions}

In this paper, we have proposed a scalable and precise application-centered call~graph generation approach \tool for Python. 
Our evaluation has demonstrated that \tool is efficient and effective. 
We have released the source code at our replication site \url{https://pythonjarvis.github.io/} with all of our experimental data set. We plan to apply \tool to foster downstream applications in security analysis and debloating Python dependencies. 
